\documentclass[aps,pra,reprint,superscriptaddress,showpacs]{revtex4-1}
\usepackage{graphicx}
\usepackage{epsfig}
\usepackage{epstopdf}
\usepackage{amsmath}
\usepackage[colorlinks=true, citecolor=blue, linkcolor=blue, urlcolor=blue]{hyperref}

\begin{document}

\title{Density dependence of the Ionization Avalanche in ultracold Rydberg gases}


\author{M.~Siercke}
\affiliation{Division of Physics and Applied Physics, Nanyang Technological
University,  21 Nanyang Link, Singapore 637371, Singapore}
\affiliation{Centre for Quantum Technologies, National University of
Singapore, 3 Science Drive 2, Singapore 117543, Singapore}
\author{F.E.~Oon}
\affiliation{Division of Physics and Applied Physics, Nanyang Technological
University,  21 Nanyang Link, Singapore 637371, Singapore}
\affiliation{Centre for Quantum Technologies, National University of
Singapore, 3 Science Drive 2, Singapore 117543, Singapore}
\author{A.~Mohan}
\affiliation{Division of Physics and Applied Physics, Nanyang Technological
University,  21 Nanyang Link, Singapore 637371, Singapore}
\author{Z.W.~Wang}
\affiliation{Division of Physics and Applied Physics, Nanyang Technological
University,  21 Nanyang Link, Singapore 637371, Singapore}
\author{M.J.~Lim}
\affiliation{Department of Physics, Rowan University, 201 Mullica Hill
Road, Glassboro, NJ 08028, USA}

\author{R.~Dumke}
\email{rdumke@ntu.edu.sg}
\affiliation{Division of Physics and Applied Physics, Nanyang Technological
University,  21 Nanyang Link, Singapore 637371, Singapore}
\affiliation{Centre for Quantum Technologies, National University of
Singapore, 3 Science Drive 2, Singapore 117543, Singapore}



\date{\today}

\begin{abstract}
We report on the behaviour of the ionization avalanche in an ensemble of ultracold $^{87}\mathrm{Rb}$ atoms coupled to a high lying Rydberg state and investigate extensions to the current model by including the effects of three-body recombination and plasma expansion. To separate the two effects we study the time dependence of the plasma formation at various densities as well as for different $nS$ and $nD$ states. At medium densities and low $n$ we observe the onset of the avalanche as has been reported in other experiments, as well as a subsequent turn-off of the avalanche for longer excitation times, which we associate with plasma expansion. At higher densities and for higher lying Rydberg states we observe a disappearance of the avalanche signature, which we attribute to three-body recombination.
\end{abstract}


\maketitle


In recent years ensembles of ultracold Rydberg atoms have become a system of interest to study due to their long range interactions and the many tools with which they can be manipulated. Electromagnetically Induced Transparency (EIT) involving Rydberg states has been used to produce strong nonlinear effects \cite{singlephoton,PRL105_193603} and the strong van der Waals interaction between Rydberg atoms has spurred considerable effort in utilizing these atoms for quantum computation and simulation protocols \cite{quantumcomp,quantumcomp2,quantumsimu}. The interesting properties of Rydberg atoms stem from their weakly bound valence electrons, giving rise to a large polarizability \cite{polarizability}. As such, they are exceptional tools for measuring and coupling to weak electric fields \cite{electricfield,adsorbate,adsorbate2,PRL108_063004}. An ultracold cloud of Rydberg atoms could also be coupled to superconducting charged or even single electron chip devices via the coulomb interaction. So far, experiments have been restricted to using the magnetic interaction between atoms and superconducting chip structures \cite{vortextrap,paper:zimmermanncoherence}. The weak bond between the electron and the nucleus is however responsible for another property of Rydberg atoms: they are easily ionized \cite{PRL101_205005}. As such, understanding the dynamics and conditions for ionization is of importance. While many aspects of the detailed physics of ionization and the formation of ultracold plasmas from Rydberg gases have been explored both experimentally and theoretically, the details of three body recombination still need further investigation \cite{PhysRep_449_77}.

Ionization of the atoms due to black body radiation or collisions can present a significant barrier to such experiments. Moreover, owing to the low velocities of the gas, ions formed in this way accumulate in the trapping region \cite{plasmaformation}, until, at a critical ion number, avalanche ionization can occur, converting all remaining Rydberg atoms into ions as nicely shown in \cite{weidemuller,PRL85_4466}. The exact dynamics of how a gas of Rydberg atoms evolves into a plasma are extremely complex and, generally, can be accurately described only using Monte Carlo methods \cite{threebody,montecarlo}. In this paper we investigate the behaviour of avalanche ionization for different $n$-states, showing surprisingly good agreement with a simple qualitative model. The employed rate equations provide insight into the fundamental roles of plasma expansion and three-body recombination (TBR) as the Rydberg gas transitions to a plasma.

We fit our data to the model used in \cite{weidemuller}, with additional terms for TBR and changing ion density:

Rubidium atoms in their $5S$ and $5P$ state ($N_g$) are excited into the Rydberg state at a rate $A$, while Rydberg atoms can decay back into the $5S$ or $5P$ states at a combined rate $B$ (eq. \ref{eq1}).

\begin{equation}
\frac{dN_g}{dt}=-A\phi N_g+BN_r\label{eq1}
\end{equation}

Here the fraction of atoms that can be excited to the Rydberg state is reduced to $\phi=exp\left(\frac{-8\eta+9\eta^2-3\eta^3}{(1-\eta)^3}\right)$ \cite{phi} and $\eta=\pi N_rR_c^3/6V$ with $N_r$ the number of Rydberg atoms, $R_c$ the Rydberg radius and $V$ the volume of the excitation region. The evolution of the number Rydberg atoms is governed by the excitation from the $5S$ and $5P$ levels, as well as ionization from black body radiation (at rate $\gamma_{bbi}$) and collisions with non-Rydberg atoms (at a rate of $\gamma_{col}=\sigma_{col}\sqrt{\frac{16k_BT}{\pi m_{Rb}}}$) \cite{penning}.

\begin{align}
\frac{dN_r}{dt}&=A\phi N_g-BN_r-\gamma_{bbi}N_r \notag \\
& \hspace{1em} -\gamma_{col}N_rN_g/V-\gamma_{av}N_rN_e/V \notag\\
& \hspace{1em} \boxed{+\gamma_{rec}N_e^2N_i/V^2} \label{eq2}
\end{align}

Due to the low initial temperature of the ground state atoms, the ions created from collisions remain in the trapping region while the hot electrons escape, until the ionic potential well becomes deep enough to retain any subsequent electrons \cite{plasmaformation}. Once such a critical number of ions is reached there is a buildup of energetic, trapped electrons $N_e$. $\gamma_{av}=\sigma_{geo}\sqrt{E_e/m_e}$ describes the rate at which collisions between the captured electrons and Rydberg atoms produce additional ions (and more electrons) where $E_e$ and $m_e$ are the average energy and the mass of the electrons, respectively. $\sigma_{geo}\approx\pi a_0 n^{\ast4}$ is the geometrical cross-section of the Rydberg atoms, with $a_0$ the Bohr radius and $n^\ast$ the effective principal quantum number. Due to its self-seeding nature and its threshold behaviour, this phenomenon is referred to as ``avalanche ionization''. The last (boxed) term of equation \ref{eq2} models the recombination of an ion back into a Rydberg atom at a rate of $\gamma_{rec}$. Since recombination requires collisions between an ion and two electrons, this process depends nonlinearly on the captured electron density \cite{threebody}. In general the ions do not recombine into their original Rydberg state, favouring instead a range of high $n$-states depending on the electron temperature \cite{threebody}. Nonetheless, the included term captures the general physics of the process: As the density of the cloud increases, TBR becomes more prominent due to its nonlinear character. A strict treatment of the state dependent recombination would require a model of the electron temperature. In our model we simplify the recombination which still gives a qualitative agreement with expected theoretical rates.

To determine the number of trapped electrons we model the ion number $N_i$ as
\begin{align}
N_i&=N_{tot}-N_g-N_r-\boxed{N_{loss}} \label{ions}\\
\frac{dN_{loss}}{dt}&=\gamma_{loss}N_i \\
N_e&=N_i-N_{crit},\hspace{1em}\mathrm{if>0},
\end{align}
with $N_{tot}$ the total number of atoms in the excitation volume. The last (boxed) term in equation \ref{ions} has been added to the model in \cite{weidemuller} in order to simulate the expansion of the plasma as a source of loss of ions from the excitation region. The exact expansion dynamics are neglected and simply modeled by a loss rate $\gamma_{loss}$. In such a way we capture the essential effects of the expansion on the system: A lowering of the ion and electron densities and an increase in the critical ion number needed to recapture electrons given by \cite{weidemuller}
\begin{equation}
N_{crit}=\frac{8E_eL\pi R^2\epsilon_0}{q^2[L(-L+\sqrt{L^2+4R^2})+4R^2\mathrm{csch}^{-1}(2R/L)]}.
\label{critical}
\end{equation}
The excitation volume is assumed to be a cylinder of radius $R$ and length $L$ as determined by the laser beams used in the Rydberg creation and $q$ is the elementary charge. The critical number further depends on the average electron energy $E_e$.4. The TBR term and the plasma expansion account for the disappearance of the avalanche at high densities as observed.

\begin{figure}[!ht]
 \includegraphics[width=\columnwidth]{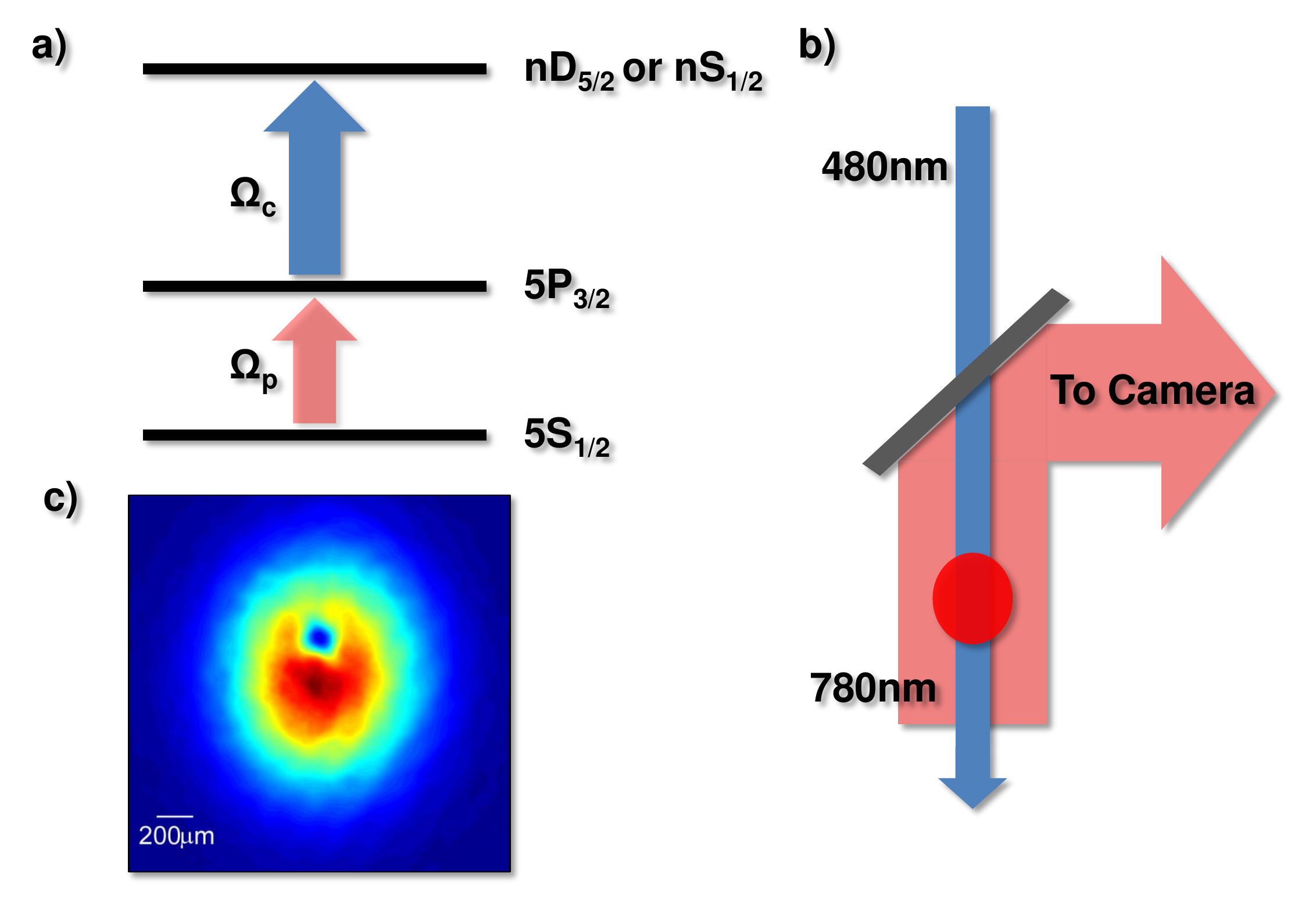}
   \caption{(Color online) Experimental setup: a) Atomic levels and laser beams involved in producing atoms in $nS$ or $nD$ Rydberg states. The probe beam ($\Omega_p$) resonantly couples the D2 transition of $^{87}Rb$ while the coupling beam ($\Omega_c$) drives the $5P_{3/2}$ level to a Rydberg state. b) The large $780\mathrm{nm}$ probe beam is used as an imaging beam while the counter-propagating $480\mathrm{nm}$ coupling beam is focused onto the atom cloud. c) At intermediate densities a hole in the cloud is observed at the location of the coupling beam when measuring the transmission of the probe beam through the atomic cloud, with an excitation time of $400\mathrm{\mu s}$.}
\label{setup}
\end{figure}

\begin{figure}[h]
 \includegraphics[width=\columnwidth]{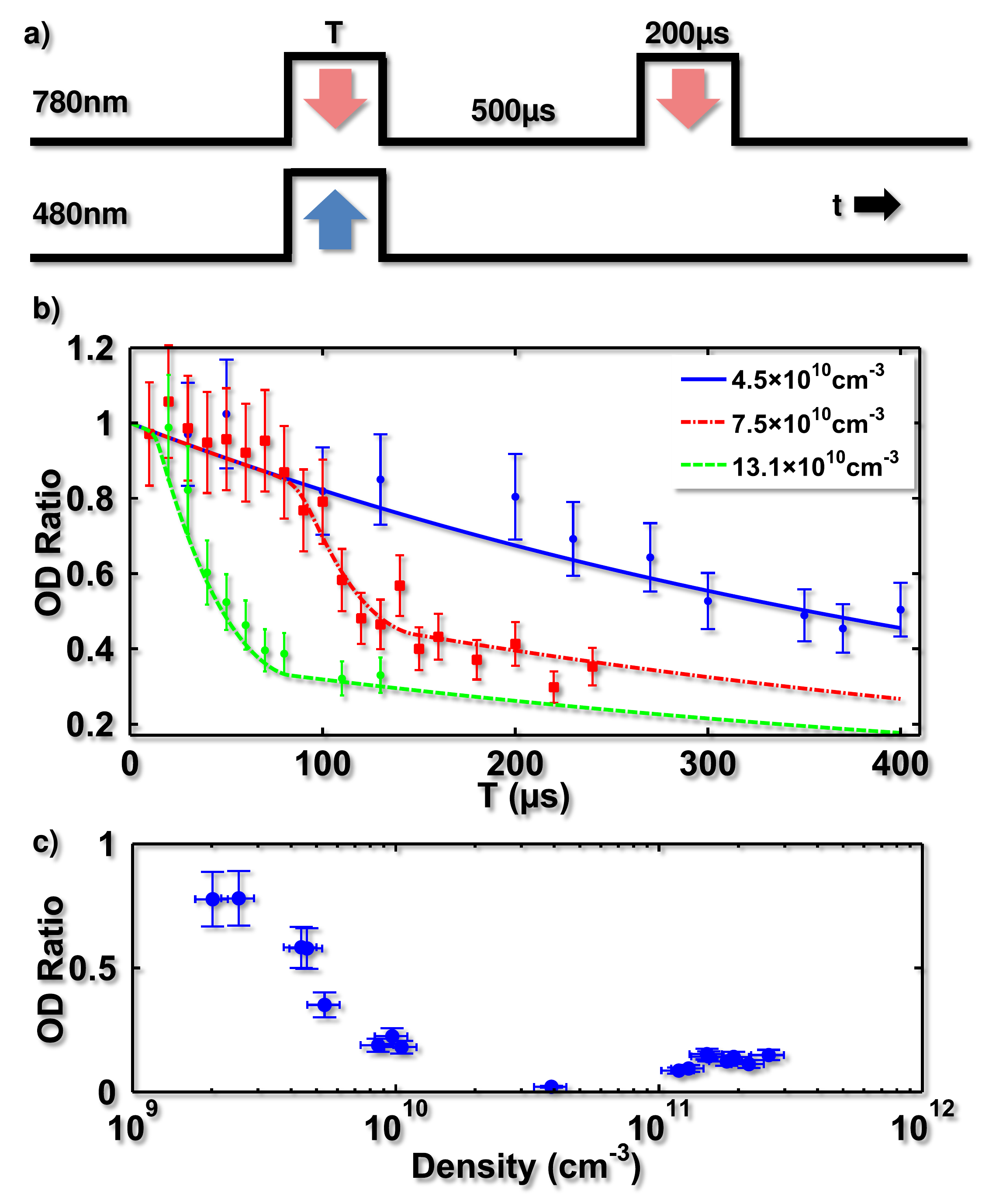}
   \caption{(Color online) a) Pulse sequence to investigate the origin and time dynamics for the hole in Optical Density formed during doubly-resonant excitation. After an initial illumination of the atoms by both coupling and probe beams for time T, absorption imaging is performed for $200\mathrm{\mu s}$ without the coupling beam. b) ODR versus time for various atomic densities. The lowering of the ODR is caused by the formation of ions from Rydberg atoms and processes not included in the model, such as atoms scattering into long-lived Rydberg states. The low and medium density data display an ionization avalanche while for low starting density the avalanche is absent due to the ion formation rate being slower than their loss rate. Fits to the data follow the model of \cite{weidemuller} with the addition of a loss term for ions and the total atom number. The loss also reproduces the turn-off of the avalanche in the low ODR region of the curves. c) The ODR (averaged over 400us of excitation) for the 40D state shows an onset of the avalanche for higher densities due to a decrease in the avalanche onset timeas seen in \ref{holeform}b and \cite{weidemuller}}.
\label{holeform}
\end{figure}

To experimentally access the different dynamic regimes of the model we produce clouds of ultracold $\mathrm{^{87}Rb}$ atoms with densities between $10^9-10^{12} \mathrm{atoms/cm^3}$ by trapping them in either a quadrupole trap or a crossed dipole trap. Before exciting atoms to a Rydberg state the trapping potential is switched off. Atoms are excited by two laser beams driving the $5S_{1/2} \rightarrow 5P_{3/2}$ and $5P_{3/2} \rightarrow nS$ or $nD$ transitions, with respective wavelengths of $780\mathrm{nm}$ and $480\mathrm{nm}$ (figure \ref{setup}).
The local density of the cloud is determined by absorption imaging. The $480\mathrm{nm}$ blue coupling beam is counter-propagating with the $780\mathrm{nm}$ red probe, which excites ground state atoms and serves as the imaging beam (figure \ref{setup}b). To separate the effects of ion loss due to plasma expansion from those due to TBR, we first investigate the time evolution of the cloud at low densities. In a typical experimental sequence the atoms are illuminated first with the coupling and probe beams and then detected after $500\mu s$ by absorption imaging (figure \ref{holeform}a). Figure \ref{setup}c shows a typical absorption image. The density distribution of the atoms has a hole in the central region, where the coupling beam was propagating. This delayed atomic absorption image implies an absence of atoms in the ground state. Considering the low power of the red beam and the short ($35\mathrm{\mu s}$) lifetime of the $40D$ state that is excited, the hole is indicative of ion formation rather than Rydberg atoms. Indeed, at the Rabi frequencies used, the three-level Bloch equations predict only $8\%$ of the atoms to be excited to the Rydberg state. Figure \ref{holeform}b shows the time dependence of the hole formation probed by varying the excitation pulse time (T). Plotted in the figure is the optical density ratio (ODR) at the central position of the blue laser beam for various atomic densities. The ODR is defined as the optical density with, divided by the optical density without the coupling beam present during the first pulse. Least square fits to the data are performed by numerically integrating equations \ref{eq1}-\ref{critical}, with $A$, $\sigma_{col}$, $\gamma_{loss}$ and $E_e$ as free parameters. $B$ is fixed, like in \cite{weidemuller}, by the steady-state solution of the optical Bloch equations. The excitation radius $R$ is chosen to be $100\mu m$, the blue beam diameter, and L is taken to be 4 times the standard deviation of the atomic cloud. All points in figure \ref{holeform}b are fit using the same parameters. For low density the figure shows a gradual decrease in ODR due to ionization from the black body radiation and collision terms in equation \ref{eq2}. Importantly, many additional processes are not accounted for in this simple model, such as recombination into long lived Rydberg states and changing electron and ion temperatures over the duration of the excitation. Based on the low density behaviour in figure \ref{holeform}b, we add an exponential decay term for the total number of atoms in the model to empirically take into account these processes. For measurements on samples of higher initial ground-state atom density, this gradual decrease in ODR is overtaken after an onset time by a faster decay. Increasing the initial ground state density further shifts the onset of the rapid decay in ODR to earlier times.

This fast decay is a sign of the ionization avalanche, with the onset given by the time required to form the critical ion number necessary to retain additional electrons (eq. \ref{critical}). The onset for avalanche depends on the density because the responsible collision rates are, themselves, functions of density. The model in \cite{weidemuller} was able to predict the early time behaviour of the Rydberg and ion population with good accuracy but deviated strongly from the data at longer timescales where the plasma dynamics may need to be taken into account. In our case, due to lower excitation rates, the production of ions is slower than in \cite{weidemuller}. While for the medium and high-density data we see the onset of the avalanche, for the low-density data we see a homogeneous formation of the hole even after the critical ion number is theoretically reached. Furthermore, at all three densities the avalanche ionization slows down at low atom numbers (long times). This behaviour is the result of the last term in equation \ref{ions}, modeling the plasma expansion as a loss of ions, as well as an extra decay on the total number of atoms. In the low density regime, ion production is slow and the expansion of the plasma can prevent or terminate the avalanche. Figure \ref{holeform}c shows the ODR of the cloud versus density when exciting the 40D state. Instead of the pulse sequence used in \ref{holeform}a the data in \ref{holeform}c is computed from a $400\mathrm{\mu s}$ absorption image of the cloud with the blue beam present, effectively averaging the hole depth over the first $400\mathrm{\mu s}$ of excitation. As predicted by the model, the critical time $t_{crit}$ for the onset of the ionization avalanche is shorter for larger densities, leading to a stronger (faster) hole formation as the density is increased from $4.5\times10^{10}\mathrm{cm^{-3}}$ to $13.1\times10^{10}\mathrm{cm^{-3}}$. The data in figure \ref{holeform}c uses higher probe ($\Omega_p=2\pi\times2.18\mathrm{MHz}$) and lower coupling ($\Omega_c=2\pi\times3.2\mathrm{MHz}$) intensity than in \ref{holeform}b, resulting in an onset of the avalanche at lower densities due to increased Rydberg excitation and Penning ionization ($5P-nD$ collision) rates. At higher densities, the data in figure \ref{holeform}c displays a slight increase in ODR, indicating that other processes, such as TBR, may become important.

To study the effects of TBR on the behaviour of the avalanche we investigate the ODR of the cloud for larger densities where the effect should be more prominent. Atoms are released either from the magnetic trap (low density/$32\mathrm{\mu K}$) or after evaporative cooling in a crossed dipole trap (high density/$1.2\mathrm{\mu K}$) and simultaneously subjected to light from the 780nm and 480nm laser beams for $\mathrm{400\mu s}$ as was done in figure \ref{holeform}c. To avoid EIT, which would complicate evaluation of the data, we keep the Rabi frequencies lower than the respective laser linewidths.

\begin{figure}[h!]
 \includegraphics[width=\columnwidth]{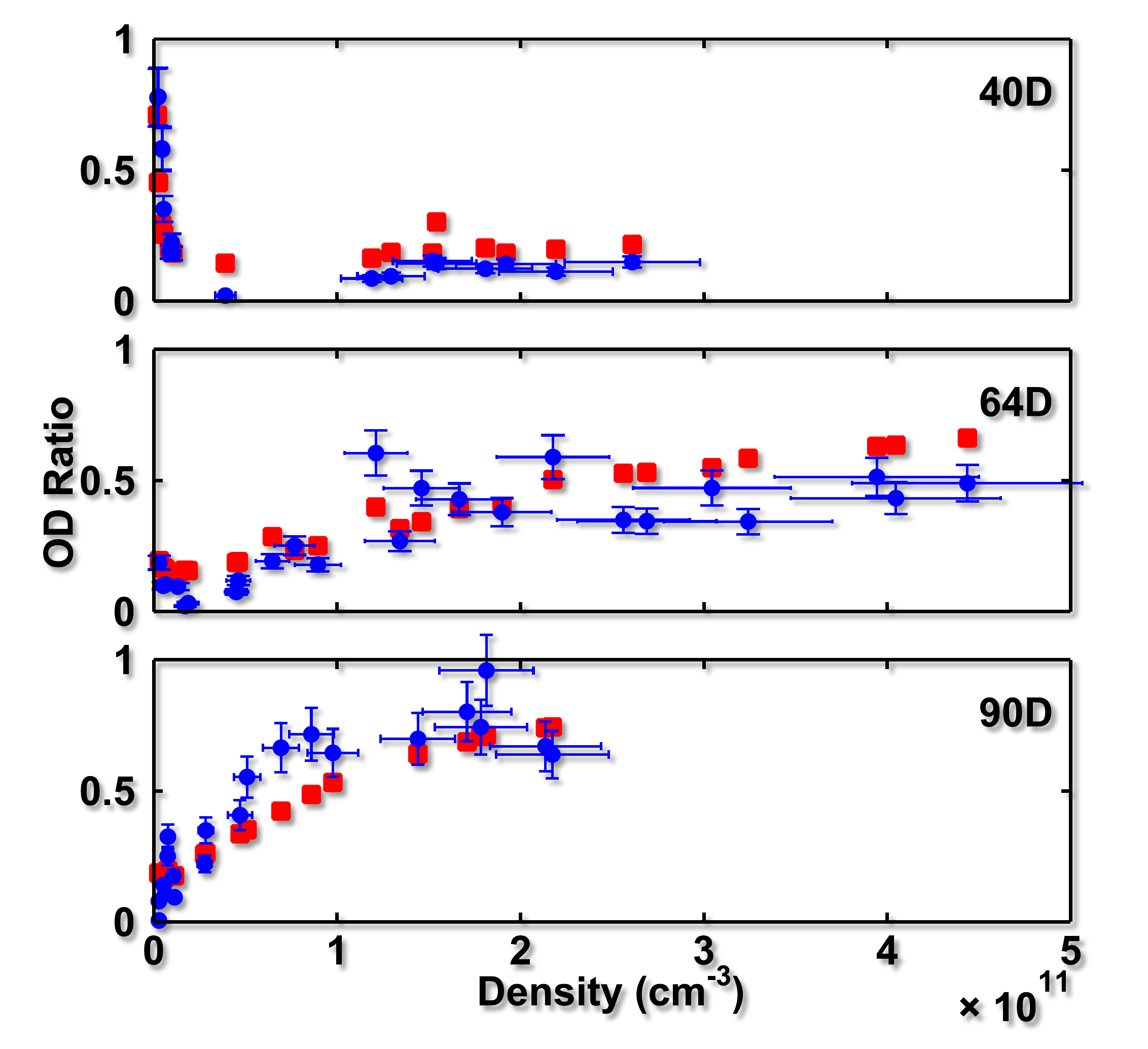}
   \caption{(Color online) ODR of different $nD$-states as a function of density (blue points). The data for the 40D state is the same as in figure \ref{holeform}c. For low density, as density and principal quantum number $n$ increase we observe a decrease in the ODR due to the ionization avalanche. Increasing the density further shows an increase in ODR, indicating that the avalanche mechanism is being suppressed. We attribute this effect to TBR, which both heats the electrons and gives the ions a short, density dependent lifetime. The red points are fits to the data, giving an electron temperature at the onset of the avalanche of $6.7\mathrm{K}$ and an electron temperature in the TBR regime of $48.7\mathrm{K}$ for the 40D state.}
\label{D-states}
\end{figure}

\begin{figure}[h!]
 \includegraphics[width=\columnwidth]{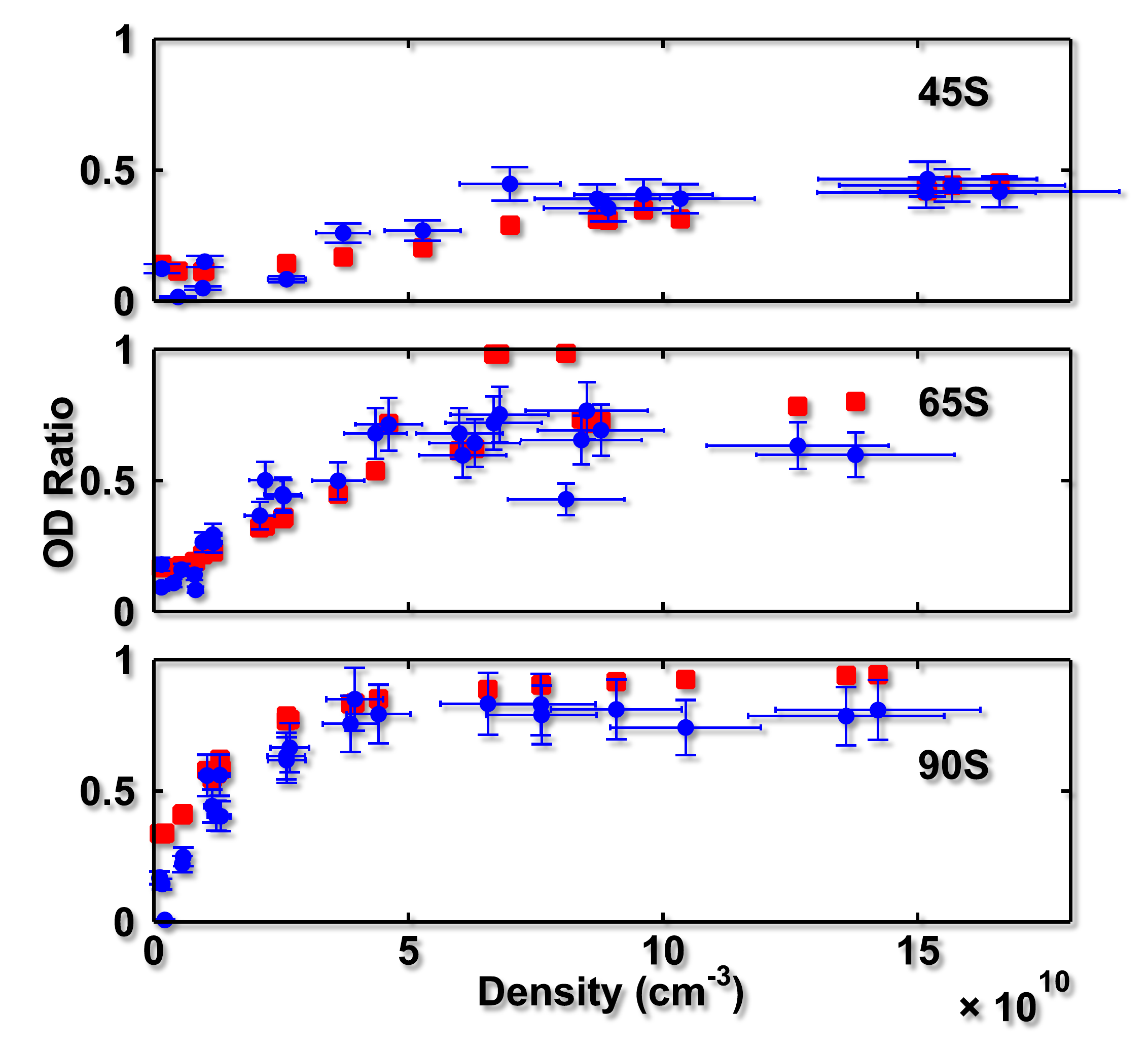}
   \caption{(Color online) ODR of different $nS$-states as a function of density (blue points). Increasing the density and principal quantum number n shows an increase in ODR, indicating that the avalanche mechanism is being suppressed. We attribute this effect to TBR, which both heats the electrons and gives the ions a short, density dependent lifetime. The red points are fits to the data, giving an electron temperature at the onset of the avalanche of $5.8\mathrm{K}$ and an electron temperature in the TBR regime of $26.0\mathrm{K}$ for the $45S$ state}
\label{S-states}
\end{figure}

Figure \ref{D-states} shows the resulting ODR as a function of density for different $nD$-states. The power of the blue beam is adjusted to produce a Rabi frequency $\Omega_c=2\pi \times 3.2\mathrm{MHz}$ for each state. The Rabi frequency of the $5S$ to $5P$ transition is chosen to be $\Omega_p=2\pi \times 2.2\mathrm{MHz}$. The data for the 40D state is the same as in figure \ref{holeform}c. For low initial densities the hole disappears for the 40D state due to the slower ionization rate, resulting in too few ions being created to overcome the ion loss. The 90D state does not show an obvious reduction in ODR at lower densities. The corresponding curves for the S-states are shown in figure \ref{S-states} with $\Omega_c=2\pi \times 6.1\mathrm{MHz}$ and $\Omega_p=2\pi \times 1.8\mathrm{MHz}$. For both D and S-states the hole disappears at high densities.

The suppression of ion production at high densities is a consequence of the TBR term in equation \ref{eq2}. As the density of the atoms increases, this rate becomes larger than the ion production rate, effectively turning off the ionization mechanisms in this regime. The red points in figures \ref{D-states} and \ref{S-states} show fits using the model, calculating the expected ODR at the experimental conditions for each point. The theory closely predicts the experimental behaviour, with TBR clearly dominating the dynamics at higher densities. However, it should be noted that the effect of TBR on the plasma is to heat up the electron cloud. This heating increases the critical number of ions needed to capture the electrons, which may also quench the avalanche dynamics. For the data in figures \ref{D-states} and \ref{S-states} we cannot distinguish between these two mechanisms. The scatter in the theoretical points stems from changes in the experimental data (such as cloud temperature) that is not reflected in the axes of the plot. The fits use the same parameters for all states in the plot (including the scaling with principal quantum number). The collective data for S and D states are fitted separately. The collision rates extracted from the fit are similar for the S and D-states ($\sigma_{col}=1.0 \pi a_0^2 n^{\ast4}$ and $0.6 \pi a_0^2 n^{\ast4}$) as are the Rydberg excitation rates ($A=3.8\times10^4$ and $4.5\times10^4$) and electron temperatures during the avalanche onset ($T_e=5.8\mathrm{K}$ for $45S$ and $6.7\mathrm{K}$) for $40D$. However, the fitted recombination rates needed to reproduce the high density behaviour differ for both S and D-states \cite{threebody}: \begin{equation}
\gamma_{rec}=\frac{1\mathrm{eV}}{k_BT_e}(13.6\mathrm{eV}/2k_BT_e)^{7/2}\times2.8\times10^{-42}\mathrm{m^6s^{-1}}.\label{recomb}
\end{equation}
 To keep the model simple we allow for separate electron temperatures for the avalanche onset and the recombination rate. This is a reasonable compromise, given that TBR is expected to heat the electrons, giving rise to a density-dependent electron temperature. The electron temperature also depends on the Rydberg state that is excited, so we include an empirical scaling of $\gamma_{rec}\propto n^{\ast4}$ instead of a scaling expected from $T_e^{-9/2}\propto n^{\ast9}$ in the recombination rate. The resulting recombination electron temperatures give $T_e=26.0\mathrm{K}$ and $48.7\mathrm{K}$ for the 45S and 40D states respectively, in good agreement with other experiments. \cite{others}

It should be stressed that, while this rate equation model captures the basic physics involved in the avalanche, it is still a simplified description of the dynamics of the system. The expansion of the plasma can lead to adiabatic cooling \cite{PhysRep_449_77}, while any residual disorder in the ion distribution and TBR will heat the system. As such, the electron and ion temperatures become time-dependent and the plasma expansion is generally a non-linear process. The close agreement of our model with the experimental results therefore indicates that the most relevant physical processes have been taken into account, with reasonable values for the temperatures and rates \cite{others}. While it is clear from the fits that TBR is responsible for the suppression of the avalanche at high density, to extract a detailed understanding of the dynamics more sophisticated models or a Monte Carlo simulation need to be employed.

In conclusion, we have investigated the density-dependent effects of plasma expansion and TBR in a gas of ultracold $^{87}Rb$ excited to various $nS$ and $nD$ Rydberg states. We observe an ionization avalanche as previously seen in other experiments and find that ion loss is responsible for the low-density suppression of the avalanche, while at high densities TBR effectively turns off the electron trapping mechanism. Electron temperatures and excitation rates found are all reasonable compared to other experiments, but we find a difference in electron temperature at the onset of avalanche ionization and during TBR. This relatively simple model does not fully treat the complex dynamics of the Rydberg-plasma system, but nonetheless captures the essential physics as evidenced by close agreement with the data. Given that TBR suppresses the formation of the plasma in our system, the effects of electron heating on the calculated recombination rate may be less than in other experiments, since the re-capture mechanism that allows for electron heating is turned off by the increased TBR rate.

We acknowledge financial support from the Centre for Quantum Technologies and A-Star, Singapore.

\end{document}